%% file: ci_paper_v2.tex
\documentclass[aps,prl,twocolumn,showpacs,superscriptaddress,groupedaddress]{revtex4-1}  
\usepackage{graphicx}  
\usepackage{dcolumn}   
\usepackage{bm}        
\usepackage{amssymb}   
\usepackage{atlasphysics}
\usepackage{hyperref} 

\def\integLumiElec{1.08} 
\def\integLumiMuon{1.21}

\def\expLimitMinusElec{9.6} 
\def\expLimitMinusMuon{8.9}
\def\expLimitMinusComb{10.4}

\def\expLimitPlusElec{9.3}
\def\expLimitPlusMuon{8.6}
\def\expLimitPlusComb{10.1}

\def\obsLimitMinusElec{10.1}
\def\obsLimitMinusMuon{8.0}
\def\obsLimitMinusComb{10.2}

\def\obsLimitPlusElec{9.4}
\def\obsLimitPlusMuon{7.0}
\def\obsLimitPlusComb{8.8}

\def\bbbar{\ensuremath{b\bar{b}}}
\def\ccbar{\ensuremath{c\bar{c}}}
\def\ttbar{\ensuremath{t\bar{t}}}

\begin{document}

\widetext
\preprint{CERN-PH-EP-2011-198}
\preprint{Submitted to Physics Letters B}
CERN-PH-EP-2011-198; submitted to Physics Letters B.

\title{Search for Contact Interactions in Dilepton Events from
 $pp$ Collisions \\ at $\sqrt{s} = 7$~\TeV{} with the ATLAS Detector
}

\author{The ATLAS Collaboration}

\begin{abstract}
This Letter presents a search for contact interactions in the dielectron and dimuon 
channels using data from proton-proton collisions produced by the 
LHC at $\sqrt{s} = 7$~\TeV{} and recorded by the ATLAS detector.
The data sample, collected in 2011, corresponds to an integrated luminosity of
$\integLumiElec$ and $\integLumiMuon$ fb$^{-1}$ in the $e^+ e^-$ and $\mu^+ \mu^-$ 
channels, respectively.
No significant deviations from the standard model are observed. 
Using a Bayesian approach with a prior flat in $1/\Lambda^2$,
the following 95\% CL lower limits are placed on the energy scale of 
$\ell\ell q q$ contact interactions:
$\Lambda^- > \obsLimitMinusElec$~\TeV{} ($\Lambda^+ > \obsLimitPlusElec$~\TeV{})
in the electron channel and
$\Lambda^- > \obsLimitMinusMuon$~\TeV{} ($\Lambda^+ > \obsLimitPlusMuon$~\TeV{})
in the muon channel
for constructive (destructive) interference in the left-left isoscalar
contact interaction model.
Limits are also provided for a prior flat in $1/\Lambda^4$.
\end{abstract}

\maketitle


\section{\label{sec:intro}Introduction}

A wide range of new physics phenomena can produce
modifications to the dilepton mass spectra predicted by the standard model (SM)
such as quark/lepton compositeness, extra dimensions, and new gauge bosons. 
The predicted form of these deviations is often either a resonance 
or an excess in the number of events in the spectra at high mass. 
This Letter reports on a search for such an excess in dilepton 
events produced in proton-proton collisions at the LHC~\cite{LHC}. 
An interpretation of these data in the context of contact interactions (CI)
is presented, 
including the first limits with the ATLAS detector in the dielectron channel and 
an update of the search performed using 2010 data in the 
dimuon channel~\cite{dimuon_ATLAS_42ipb}. 
A separate paper describes the search for new heavy resonances in the dilepton mass 
spectra performed using the same ATLAS dataset~\cite{Zprime}.

If quarks and leptons are composite, with at least one common constituent, 
the interaction of these constituents would likely be manifested 
through an effective four-fermion contact interaction at energies
well below the compositeness scale. 
Such a contact interaction could also describe a new interaction with a messenger 
too heavy for direct observation at the LHC,
in analogy with Fermi's nuclear $\beta$~decay theory~\cite{Fermi}.

The Lagrangian for a general contact interaction has
the form~\cite{ELP}

\begin{equation}\label{lagrangian}
\begin{array}{r@{\,}c@{}c@{\,}l@{\,}l}
\mathcal L = \frac{g^2}{2\Lambda^2}\;[&& \eta_{LL}&\, \overline\psi_L\gamma_{\mu}\psi_L\,\overline\psi_L\gamma^{\mu}\psi_L \nonumber \\
& +&\eta_{RR}& \overline\psi_R\gamma_{\mu}\psi_R \,\overline\psi_R\gamma^{\mu}\psi_R \\
&+&2\eta_{LR}& \overline\psi_L\gamma_{\mu}\psi_L \,\overline\psi_R\gamma^{\mu}\psi_R&]\: ,\nonumber
\end{array}
\end{equation}
where $g$ is a coupling constant chosen to obey $g^2/4\pi = 1$,
$\Lambda$ is the contact interaction scale, which in the context of compositeness
models is the energy scale below which fermion constituents are bound,  
and $\psi_{L,R}$ are left-handed and right-handed fermion fields, respectively.
The parameters $\eta_{ij}$, where $i$ and $j$ are $L$ or $R$, 
define the chiral structure (left or right) of the new interaction.
Specific models are constructed by setting different combinations of these parameters
to assume values of $-1$, $0$ or $+1$.
The addition of this contact interaction term to the SM Lagrangian 
alters the Drell-Yan (DY) production cross section ($q \bar{q} \to Z/\gamma^\ast \to \ell^+\ell^-$).
The largest deviations, either constructive or destructive, are expected at high dilepton invariant 
mass and are determined by the scale $\Lambda$ and the sign of the parameter $\eta_{ij}$.
This analysis interprets the data in the context of the left-left isoscalar model (LLIM), 
which is commonly used as a benchmark for contact interaction searches~\cite{PDG}.
The LLIM is defined by setting $\eta_{LL} = \pm 1$ and $\eta_{RR}$ = $\eta_{LR}$ = 0. 

With the introduction of a contact interaction, the differential cross section
for the process $q \bar{q} \to \ell^+\ell^-$ can be written
\begin{equation}\label{cross_section}
\frac{d\sigma}{dm_{\ell\ell}} = \frac{d\sigma_{DY}}{dm_{\ell\ell}}
 - \eta_{LL}\frac{F_I(m_{\ell\ell})}{\Lambda^2} + \frac{F_C(m_{\ell\ell})}{\Lambda^4}~,
\end{equation}
where $m_{\ell\ell}$ is the final-state dilepton mass.
The expression above includes a SM DY term, as well as DY-CI interference ($F_I$) 
and pure contact interaction ($F_C$) terms
(see Ref.~\cite{EHLQ} for the full form of this expression).
At the largest $\Lambda$ values to which this analysis is sensitive,
both interference and pure contact interaction terms play a significant role.
For example, at dilepton masses greater than 300~\GeV{} and $\Lambda = 9$~\TeV{},
the magnitude of the interference term is about 1.5 times that of the
pure contact interaction term. 

The present analysis focuses on identifying a broad deviation from the SM dilepton
mass spectra, which are expected to be dominated by the DY process.
Current experimental bounds on $\Lambda$ (see below) indicate any deviation 
from a new interaction would appear at masses well above the $Z$ boson peak.
Consequently, the search region is restricted to dilepton masses above 150~\GeV{}. 
The analysis exploits the high $pp$ collision energy of the LHC
and the capabilities of the ATLAS detector to identify and reconstruct
electrons and muons at high momentum.

  Previous searches for contact interactions have been carried out 
in neutrino scattering~\cite{NuTeV},
as well as at electron-positron~\cite{ALEPH,DELPHI,DELPHI_eebb,L3,OPAL}, 
electron-proton~\cite{H1,ZEUS}, and 
hadron colliders~\cite{Abe97,Affolder01,Abulencia06,Abbott99,Abazov09,dijet_CMS_3ipb,dijet_ATLAS_3ipb,dijet_CMS_36ipb,dijet_ATLAS_36ipb}.
In the case of $e e q q$ contact interactions, the best limits in the LLIM
for all quark flavors
come from $e^+ e^-$ experiments with
$\Lambda^- > 7.2$~\TeV{} and
$\Lambda^+ > 12.9$~\TeV{}~\cite{ALEPH} at 95\% confidence level (CL) for
$\eta_{LL} = -1$ and $+1$, respectively.
These limits assume that contact interactions of electrons with all quark
flavors are of the same strength.
Best limits set in the specific case of first generation quarks are
$\Lambda^- > 9.1$~\TeV{} and
$\Lambda^+ > 8.6$~\TeV{}~\cite{OPAL} at 95\% CL.
In the case of $\mu\mu q q$ contact interactions, the best limits
are $\Lambda^- > 4.9$~\TeV{} and $\Lambda^+ > 4.5$~\TeV{}
from the ATLAS analysis of the 2010 
data~\cite{dimuon_ATLAS_42ipb}.

\section{\label{sec:atlas}ATLAS Detector and Data Sample}
ATLAS is a multipurpose particle detector~\cite{detector}.
It consists of an inner tracking detector surrounded by a 2~T superconducting 
solenoid, electromagnetic and hadronic calorimeters, and a muon spectrometer 
with a toroidal magnetic field. Charged particle tracks are reconstructed 
using the inner detector, which comprises a silicon pixel detector, 
a silicon-strip tracker, and a transition radiation tracker, covering 
the pseudorapidity range $|\eta|$ $<$ 2.5~\cite{coordinates}. 
A hermetic calorimeter, which covers $|\eta|$ $<$ 4.9, surrounds the 
superconducting solenoid. 
The liquid-argon electromagnetic calorimeter, which plays an important 
role in electron identification and measurement, 
is finely segmented, with readout granularity 
($\eta$, $\phi$) varying by layer and cells as small as $0.025 \times 0.025$ extending 
to $|\eta| < 2.5$, to provide excellent energy and position resolution.
The electron energy resolution is dominated at high energy
by a constant term equal to 1.2\% in the barrel ($|\eta| < 1.37$) 
and 1.8\% in the endcaps  ($1.52 < |\eta| < 2.47$).
Hadron calorimetry is provided by an 
iron-scintillator tile calorimeter in the central rapidity range  $|\eta| < 1.7$ 
and a liquid-argon calorimeter in the rapidity range $1.5 < |\eta| < 4.9$. 
Another key detector component for this analysis is the muon spectrometer, 
which is designed to identify muons and measure their momenta with high accuracy.
The currently achieved resolution for momenta transverse to the beam line ($p_T$) 
of 1~\TeV{} ranges from 15\% (central) to 44\% (for $|\eta| > 2$).
The muon system comprises three toroidal magnet systems,
a trigger system consisting of resistive plate chambers in the barrel and 
thin-gap chambers in the endcaps, providing triggering capability up to $|\eta| = 2.4$,
and a set of precision monitored drift tubes and cathode strip chambers 
in the region $|\eta| < 2.7$.

The data sample for this analysis was collected during LHC operation 
in the first half of 2011 and corresponds to a total integrated luminosity
of $\integLumiElec$ and $\integLumiMuon$ fb$^{-1}$
in the $e^{+}e^{-}$ and $\mu^{+}\mu^{-}$ channels, respectively. 
It was collected with stable beam conditions and an operational inner detector.
For the electron (muon) channel, the calorimeter (muon spectrometer) was also
required to be operational.
Events were selected by requiring that they pass the single electron (muon) 
trigger with a transverse momentum $p_T$ 
threshold of 20 (22) \GeV{}. This analysis follows the same event selection 
as the search for new heavy resonances. A summary is provided below,
a more complete description can be found in Ref.~\cite{Zprime}.

\section{\label{sec:sim}Signal and Background Modeling}
This analysis looks for deviations from the expected SM dilepton spectra. 
The largest SM contribution comes from DY followed by 
semileptonic decay of $\ttbar$ pairs, 
electroweak diboson production ($WW$, $WZ$, and $ZZ$), and 
production of jets in association with a $W$ boson ($W$+jets).
In addition, multi-jet production (QCD) is a significant 
background in the electron channel.
With the exception of QCD, Monte Carlo (MC) simulation was used to model these backgrounds.

DY events were generated with {\sc Pythia} 6.421~\cite{Pythia} and 
{\sc Mrst2007} LO$^\ast$ parton distribution functions (PDFs)~\cite{MRST}.
Signal DY+CI samples in the LLIM were generated with the same version of
{\sc Pythia} for the full dilepton differential cross section 
as shown in Eq.~(\ref{cross_section}).
This ensured that the interference term $F_{I}$ was properly included. 
All quark flavors contributed to the contact interaction in these signal samples.
Diboson processes were produced with {\sc Herwig} 6.510~\cite{Herwig}
using {\sc Mrst2007} LO$^\ast$ PDFs.
The $W$+jets background was generated with {\sc Alpgen}~\cite{Alpgen}
and {\sc Cteq6l1}~\cite{CTEQ6L1} PDFs,
and the $\ttbar$ background with {\sc Mc@nlo} 3.41~\cite{MCatNLO}
and {\sc Cteq6.6}~\cite{CTEQ66} PDFs. 
For the latter two, {\sc Jimmy} 4.31~\cite{Jimmy} was used 
to describe multiple parton interactions and {\sc Herwig} to describe the remaining underlying
event and parton showers. 
{\sc Photos}~\cite{Photos} was used to handle the final-state photon radiation 
for all MC samples. 
Furthermore, higher order QCD corrections were implemented
via a mass-dependent $K$-factor defined as the ratio between 
the next-to-next-to-leading order (NNLO)
$Z/ \gamma^\ast$ cross section, calculated using {\sc Phozpr}~\cite{Phozpr} and 
{\sc Mstw2008} PDFs~\cite{MSTW}, and the LO cross section.
This QCD $K$-factor was applied to both DY and DY+CI samples.
Likewise, DY and DY+CI samples are corrected with a mass-dependent
$K$-factor accounting for higher-order 
electroweak corrections arising from virtual heavy gauge boson loops 
that are calculated using {\sc Horace}~\cite{Horace}. 
Finally, the generated samples were processed through a full simulation of the
ATLAS detector~\cite{ATLASsimul} based on the {\sc Geant}~4 package~\cite{Geant}.

For both channels, the QCD multi-jet background is evaluated from data due to 
poor modeling and low MC statistics.
In the electron channel, a reversed electron identification technique is used to 
select a sample of events in which both electrons
fail a subset of the electron identification criteria (see further discussion below).
This sample is then used to determine the shape of the
QCD background as a function of dielectron invariant mass.
This template shape and the sum of the 
DY, dibosons, $\ttbar$, and $W$+jets backgrounds 
normalized by their cross sections (including higher order corrections)
are fitted to the observed dielectron mass distribution 
in the range between 70 and 200 \GeV{} to determine
the normalization of the QCD contribution.
The above QCD background estimate is cross-checked with two other methods
described in Ref.~\cite{Zprime} in order to determine its systematic
uncertainty. In particular, these cross-checks set bounds on
the potential bias in the QCD mass spectrum
introduced by the reversed identification technique.
In the muon channel, the QCD background is much smaller and is also evaluated from data. 
A reverse isolation method is utilized: a QCD sample is selected from data 
by requiring two non-isolated muons with 
$0.1 < \Sigma p_T(R < 0.3) / p_T(\mu) < 1.0$,
where $\Sigma p_T$ ($R < 0.3$) is the sum of the $p_T$ of the
tracks in a cone of $R = \sqrt{\Delta\eta^2 + \Delta\phi^2} < 0.3$ 
around the direction of the muon.
The normalization for this sample is obtained from the ratio of isolated 
to non-isolated dimuon events in QCD $\ccbar$ and $\bbbar$ MC,
where the isolation requirement is $\Sigma p_T(R < 0.3) / p_T(\mu) < 0.05$.
Muons from light hadron decays are not a significant source of background
at the high momenta relevant to this analysis.

\section{\label{sec:event}Event Selection}
Events passing the trigger selection described above are
required to have a pair of either electrons or muons
with $p_T$ greater than 25 \GeV{} to ensure maximal trigger efficiency. 
To reject cosmic ray events and beam halo background, events are required to have a 
reconstructed vertex with at least three charged particle tracks with 
$p_T > 0.4$ \GeV{}.
If several such vertices are found, the vertex with the largest $\Sigma p_T^2$
is selected as the primary vertex of the event, where the sum is over all
charged particles associated with the given vertex.
Electron candidates are confined in $|\eta| < 2.47$, with
the detector crack region 1.37 $\leq$ $|\eta|$ $\leq$ 1.52 excluded
because of degraded energy resolution.
Muon candidates are required to be within the inner detector acceptance.

Electron candidates are formed from clusters of cells reconstructed
in the electromagnetic calorimeter. Identification criteria on the transverse shower shape, 
the longitudinal leakage into the hadronic calorimeter, and the association 
to an inner detector track are applied to the cluster to satisfy 
the medium electron definition~\cite{El_med}.
The electron energy is obtained from the calorimeter measurements
and its direction from the associated inner detector track.
A hit in the first layer of the pixel detector is required (if an active
pixel layer is traversed) to suppress background from photon conversions.
Further QCD jet background suppression is achieved
by demanding the highest $p_T$ electron in the event to be
isolated. To that effect,
the sum of the calorimeter transverse momenta around the electron
direction $\Sigma p_T (R < 0.2)$ must be less than 7 \GeV{}.
The core of the electron energy deposition is excluded and the
sum is corrected for transverse shower leakage and pile-up from
additional $pp$ collisions.
In addition, the two electron candidates are not required to have opposite 
charge because of possible charge mis-identification either 
due to bremsstrahlung or to the limited momentum resolution of the inner 
detector at very high $p_T$.
If the event contains more than two selected electrons, the two
electrons with the highest $p_T$ are chosen.
For these selection criteria, the overall event acceptance for signal events
has a small dependence on the dielectron mass above 500 \GeV{} and
a value of approximately 65\% at 1~\TeV{}.

Muon candidates are required to be of opposite charge and 
are reconstructed independently in both the inner detector and the muon spectrometer.
The momentum is taken from a combined fit to the measurements from both subsystems. 
To obtain optimal momentum resolution and accurate modeling by the simulation, 
the muon candidates are required to satisfy the following requirements
in the muon spectrometer: have at least three hits in each of the inner, 
middle, and outer detectors, and at least one hit 
in the non-bending $xy$ plane.
To suppress background from cosmic rays, requirements are imposed on the
muon impact parameter and primary vertex (PV): 
transverse impact parameter $|d_0| <$ 0.2~mm, 
$z$ coordinate with respect to the PV $|z_0 - z_{\rm PV}|<$ 1~mm, 
and $z$ position of the PV $|z_{\rm PV}| <$ 200~mm.
Muons are also required to be isolated to reduce background from jets: 
$\Sigma p_T (R < 0.3) / p_T(\mu) <$ 0.05,
as explained in the previous section.
If more than one opposite-sign muon pair is found in an event,
the pair with the largest $p_T(\mu^+) + p_T(\mu^-)$ is chosen.
The overall event acceptance for signal events
has only a weak dependence on the dimuon mass with
a value of approximately 40\% at 1~\TeV{}.
Stringent requirements on the presence of hits in all three layers of the 
muon spectrometer and the limited three-layer geometrical coverage
are the primary reason for the lower acceptance relative
to the electron channel.

Extensive comparisons between data and MC simulation were performed
at the level of single-lepton distributions
to confirm that the simulation reproduces the selected data well,
especially at high momentum.

\begin{figure}
  \includegraphics[width=0.48\textwidth]{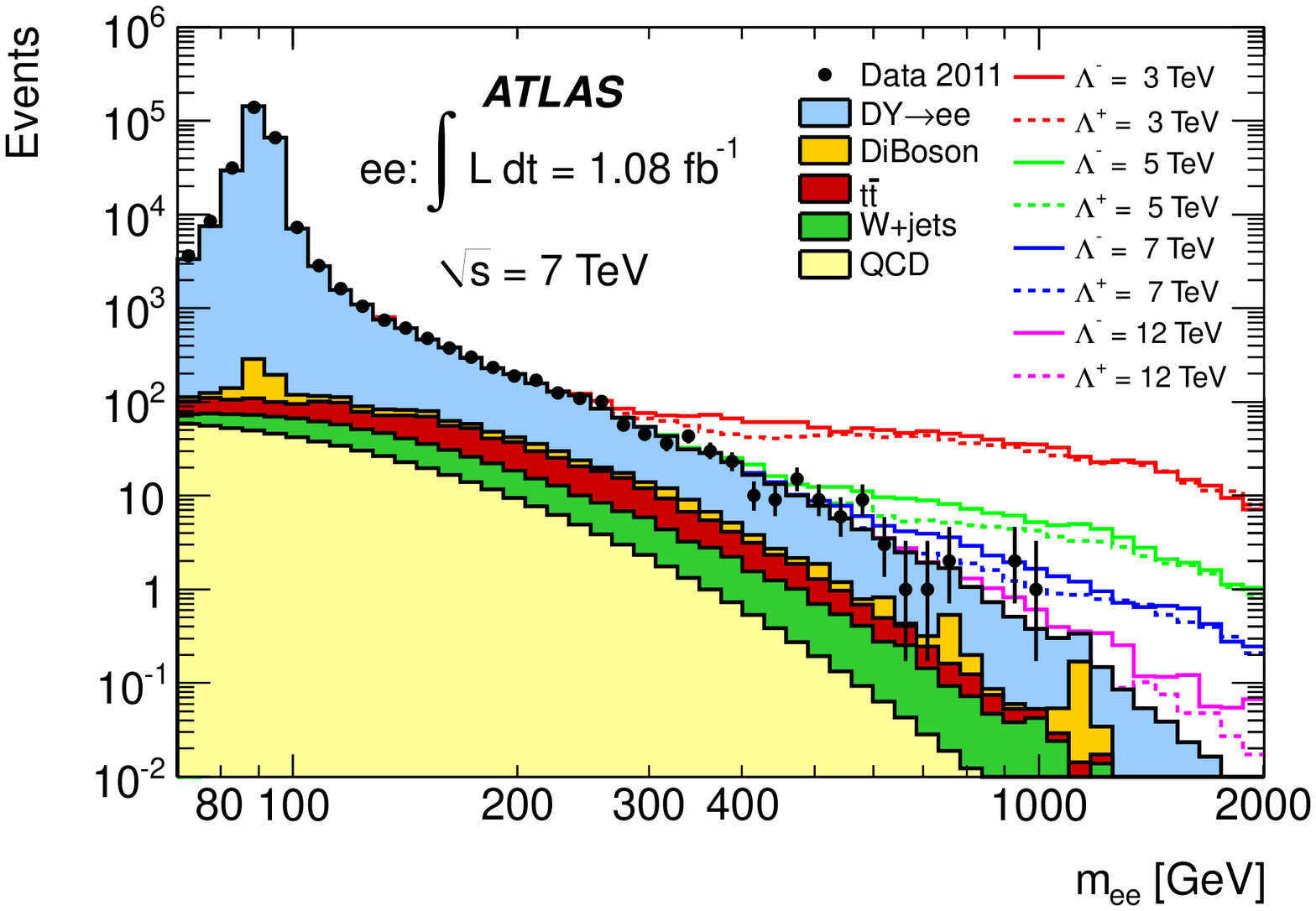}
  \includegraphics[width=0.475\textwidth]{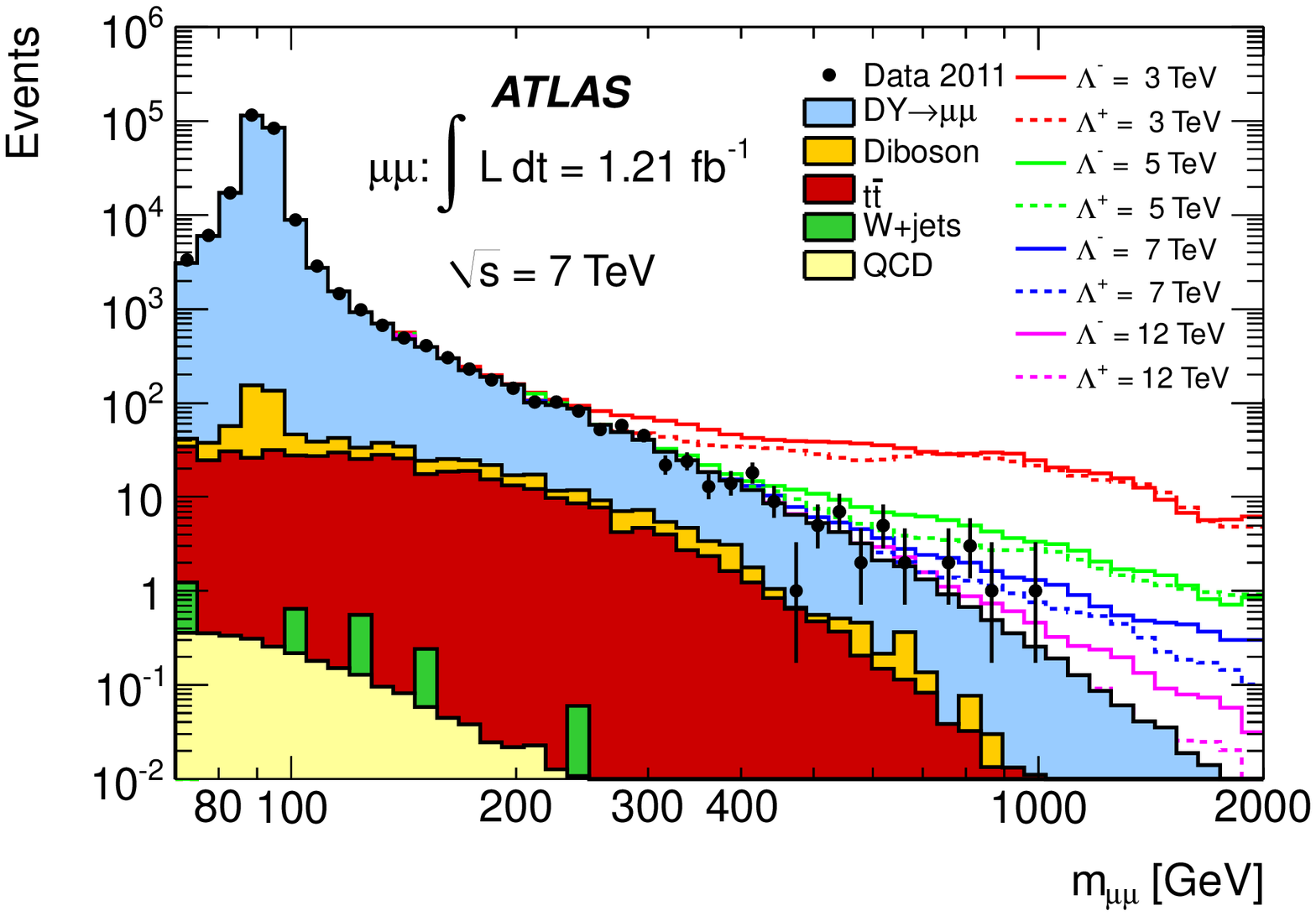}
  \caption{\label{fig:mass} Dielectron (top) and dimuon (bottom) invariant mass distributions
  for data (points) and Monte Carlo simulation (histograms). The open histograms correspond to
  the distributions expected in the presence of contact interactions with different values
  of $\Lambda$ for both constructive (solid histograms) and destructive (dashed histograms)
  interference.}
\end{figure}

Figure~\ref{fig:mass} displays the dielectron and
dimuon mass spectra for all selected events with 
invariant mass greater than 70~\GeV{}. 
The expected event yields for the different processes are obtained
by first normalizing each MC process by its cross section
(including higher order corrections)
and then normalizing the total MC event yield plus the data-derived QCD background
to the data in the $Z$ peak region (dilepton mass between 70 and 110~\GeV{}).
Good agreement is observed between the data and the SM prediction over the whole
dilepton mass range. A quantitative comparison is provided in Tables~\ref{dielec_yields}
and \ref{dimuon_yields}. 
A slight excess of events observed at high dimuon mass is consistent 
with a statistical fluctuation.
The most significant deviation in the number of dimuon events occurs 
for events with mass greater than 800~\GeV{}. 
In this region, the Poisson probability for observing 5 or more
events where 2.1 are expected is 6.2\%.
The muon tracks in the five data events were inspected in detail and no
problem was found.

\begin{table*}[tb]
\newcommand\T{\rule{0pt}{2.6ex}}
\newcommand\B{\rule[-1.2ex]{0pt}{0pt}}
\caption{Expected and observed numbers of events in the electron channel.
 The total expected yield is normalized to the data in the $Z$ peak control region
 between 70 and 110~\GeV{}.
 The errors quoted originate from both systematic uncertainties and limited MC statistics,
 except the error on the total expected yield in the normalization region
 which is given by the square root of the number of observed events.
}
\label{dielec_yields}
\centering
\scriptsize
\begin{tabular}{lccccccc}
\hline
\hline
$m_{ee}$ [\GeV{}] \T \B & \multicolumn{1}{c}{70-110} & \multicolumn{1}{c}{110-130} & \multicolumn{1}{c}{130-150} & \multicolumn{1}{c}{150-170} & \multicolumn{1}{c}{170-200} & \multicolumn{1}{c}{200-240} & \multicolumn{1}{c}{240-300} \\
\hline
DY \T & 258482 $\pm$ 410 & 3185 $\pm$ 110 & 1183 $\pm$ 46 & 608 $\pm$ 28 & 473 $\pm$ 24 & 312 $\pm$ 18 & 196 $\pm$ 9 \\
\ttbar \T & 218 $\pm$ 36 & 87 $\pm$ 7 & 64 $\pm$ 5 & 51 $\pm$ 2 & 51 $\pm$ 2 & 37 $\pm$ 2 & 30 $\pm$ 1 \\
Dibosons \T & 368 $\pm$ 19 & 31 $\pm$ 2 & 24 $\pm$ 2 & 15 $\pm$ 1 & 16 $\pm$ 1 & 14 $\pm$ 1 & 8 $\pm$ 1 \\
$W$+jets \T & 150 $\pm$ 100 & 57 $\pm$ 11 & 40 $\pm$ 9 & 27 $\pm$ 6 & 26 $\pm$ 6 & 20 $\pm$ 5 & 14 $\pm$ 4 \\
QCD \T & 332 $\pm$ 60 & 80 $\pm$ 42 & 50 $\pm$ 20 & 32 $\pm$ 7 & 29 $\pm$ 7 & 19 $\pm$ 15 & 12 $\pm$ 9 \\
\hline
 Total \T \B & 259550 $\pm$ 510 & 3440 $\pm$ 120 & 1361 $\pm$ 50 & 733 $\pm$ 30 & 595 $\pm$ 25 & 401 $\pm$ 24 & 260 $\pm$ 13 \\
\hline
Data \T \B & \multicolumn{1}{c}{259550}  &  \multicolumn{1}{c}{3419}  &  \multicolumn{1}{c}{1362}  &  \multicolumn{1}{c}{758}  &  \multicolumn{1}{c}{578}  &  \multicolumn{1}{c}{405}  &  \multicolumn{1}{c}{256} \\
\hline
\hline
 \multicolumn{8}{c}{ }  \\
\hline
\hline
\T $m_{ee}$ [\GeV{}] \T \B & \multicolumn{1}{c}{300-400} & \multicolumn{1}{c}{400-550} & \multicolumn{1}{c}{550-800} & \multicolumn{1}{c}{800-1200} & \multicolumn{1}{c}{1200-1800} & \multicolumn{1}{c}{1800-3000} \\
\hline
DY \T & 105.0 $\pm$ 5.0 & 41.0 $\pm$ 2.2 & 12.8 $\pm$ 0.8 & 2.5 $\pm$ 0.2 & 0.29 $\pm$ 0.05 & $<0.05$ \\
\ttbar \T & 14.9 $\pm$ 0.8 & 4.5 $\pm$ 0.2 & 1.0 $\pm$ 0.1 & 0.10 $\pm$ 0.02 & $<0.05$ & $<0.05$ \\
Dibosons \T & 7.5 $\pm$ 1.1 & 2.1 $\pm$ 0.4 & 1.0 $\pm$ 0.3 & 0.3 $\pm$ 0.1 & $<0.05$ & $<0.05$ \\
$W$+jets \T & 9.0 $\pm$ 3.2 & 3.5 $\pm$ 1.6 & 1.0 $\pm$ 0.7 & 0.2 $\pm$ 0.3 & $<0.05$ & $<0.05$ \\
QCD \T & 5.5 $\pm$ 4.4 & 1.5 $\pm$ 1.2 & 0.3 $\pm$ 0.2 & $<0.05$ & $<0.05$ & $<0.05$ \\
\hline
 Total \T \B & 141.9 $\pm$ 8.0 & 52.6 $\pm$ 3.0 & 16.1 $\pm$ 1.1 & 3.0 $\pm$ 0.4 & 0.33 $\pm$ 0.05 & $<0.05$ \\
\hline
Data \T \B & \multicolumn{1}{c}{147}  &  \multicolumn{1}{c}{48}  &  \multicolumn{1}{c}{17}  &  \multicolumn{1}{c}{3}  &  \multicolumn{1}{c}{0}  &  \multicolumn{1}{c}{0} \\
\hline
\hline
\end{tabular}
\end{table*}

\begin{table*}[tb]
\newcommand\T{\rule{0pt}{2.6ex}}
\newcommand\B{\rule[-1.2ex]{0pt}{0pt}}
\caption{Expected and observed numbers of events in the dimuon channel.
 The total expected yield is normalized to the data in the $Z$ peak control
 region between 70 and 110~\GeV{}.
 The errors quoted originate from both systematic uncertainties and limited MC statistics,
 except the error on the total expected yield in the normalization region
 which is given by the square root of the number of observed events.
}
\label{dimuon_yields}
\centering
\scriptsize
\begin{tabular}{lccccccc}
\hline
\hline
$m_{\mu\mu}$ [\GeV{}] \T \B & \multicolumn{1}{c}{70-110} & \multicolumn{1}{c}{110-130} & \multicolumn{1}{c}{130-150} & \multicolumn{1}{c}{150-170} & \multicolumn{1}{c}{170-200} & \multicolumn{1}{c}{200-240} & \multicolumn{1}{c}{240-300} \\
\hline
DY \T    &  236405 $\pm$ 320  &  3133 $\pm$ 90  &  1076 $\pm$ 36  &  548 $\pm$ 22  &  417 $\pm$ 18  &  249 $\pm$ 13  &  153 $\pm$ 7  \\ 
\ttbar \T  &  193 $\pm$ 21  &  70 $\pm$ 9  &  51 $\pm$ 7  &  34 $\pm$ 4  &  38 $\pm$ 4  &  30 $\pm$ 3  &  21 $\pm$ 2  \\ 
Diboson \T &  307 $\pm$ 16  &  25 $\pm$ 2  &  19 $\pm$ 2  &  13 $\pm$ 2  &  12 $\pm$ 1  &  10 $\pm$ 1  &  8 $\pm$ 1  \\ 
$W$+jets \T &  1 $\pm$ 1  &  $<0.5$  &  $<0.5$  &  $<0.5$  &  $<0.5$  &  $<0.5$  &  $<0.5$  \\ 
QCD   \T  &  1 $\pm$ 1  &  $<0.5$  &  $<0.5$  &  $<0.5$  &  $<0.5$  &  $<0.5$  &  $<0.5$ \\
\hline
Total  \T \B &  236908 $\pm$ 490  &  3229 $\pm$ 90  &  1147 $\pm$ 37  &  595 $\pm$ 22  &  467 $\pm$ 19  &  290 $\pm$ 13  &  182 $\pm$ 8  \\
\hline
Data \T \B & \multicolumn{1}{c}{236908}  &  \multicolumn{1}{c}{3211}  &  \multicolumn{1}{c}{1132}  &  \multicolumn{1}{c}{621}  &  \multicolumn{1}{c}{443}  &  \multicolumn{1}{c}{279}  &  \multicolumn{1}{c}{195} \\
\hline
\hline
 \multicolumn{8}{c}{ }  \\
\hline
\hline
\T $m_{\mu\mu}$ [\GeV{}] \T \B & \multicolumn{1}{c}{300-400} & \multicolumn{1}{c}{400-550} & \multicolumn{1}{c}{550-800} & \multicolumn{1}{c}{800-1200} & \multicolumn{1}{c}{1200-1800} & \multicolumn{1}{c}{1800-3000} \\
\hline
DY    \T  &  80.8 $\pm$ 3.9  &  31.0 $\pm$ 1.7  &  9.2 $\pm$ 0.6  &  1.8 $\pm$ 0.2  &  0.22 $\pm$ 0.04  &  $<0.05$ \\ 
\ttbar \T  &  11.7 $\pm$ 1.2  &  3.5 $\pm$ 0.3  &  0.7 $\pm$ 0.1  &  0.06 $\pm$ 0.02 &  $<0.05$  &  $<0.05$ \\ 
Diboson \T &  6.7 $\pm$ 1.1  &  1.0 $\pm$ 0.4  &  0.7 $\pm$ 0.3  &  $<0.05$  &  $<0.05$  &  $<0.05$ \\ 
$W$+jets \T &  $<0.05$  &  $<0.05$  &  $<0.05$  &  $<0.05$  &  $<0.05$  &  $<0.05$ \\ 
QCD    \T &  $<0.05$  &  $<0.05$  &  $<0.05$  &  $<0.05$  &  $<0.05$  &  $<0.05$ \\ 
\hline
Total \T \B  &  99.3 $\pm$ 4.2  &  35.5 $\pm$ 1.8  &  10.6 $\pm$ 0.7  &  1.9 $\pm$ 0.2  &  0.22 $\pm$ 0.04  &  $<0.05$  \\ 
\hline
Data \T \B & \multicolumn{1}{c}{83}  &  \multicolumn{1}{c}{39}  &  \multicolumn{1}{c}{12}  &  \multicolumn{1}{c}{5}  &  \multicolumn{1}{c}{0}  &  \multicolumn{1}{c}{0} \\
\hline
\hline
\end{tabular}
\end{table*}

\section{\label{sec:systematics}Systematic Uncertainties}
Since the MC event yields are normalized to the number of events observed
in the $Z$ peak region, only mass-dependent systematic uncertainties
need to be considered,
except for a 5\% overall uncertainty in the knowledge of
the $Z/\gamma^\ast$ cross section in the normalization region.
This overall uncertainty is required since the cross section change due to 
the new physics is defined with respect to the SM cross section.
The dominant uncertainties are of theoretical origin but experimental sources
are also considered. 

Theoretical uncertainties in the predicted event yields
arise from the limited knowledge of
PDFs, $\alpha_S$, QCD and electroweak $K$-factors, 
and $Z/\gamma^\ast$ cross section. 
The finite available MC statistics are also taken into account.
The uncertainty due to the PDF and $\alpha_S$ is estimated using the {\sc Mstw2008}
PDF eigenvector set and additional PDFs corresponding to variations in $\alpha_S$.
The resulting effect is about 4\% at the $Z$ pole growing with increasing mass
to 10\% at 1.5~\TeV{}.
Uncertainties due to QCD and electroweak $K$-factors are estimated to grow
from 0.3\% and 0.4\% at the $Z$ pole to
3\% and 4.5\% at 1.5~\TeV{}, respectively, for both electron and muon channels.
Estimates for the QCD $K$-factor are obtained by varying the renormalization
and factorization scales independently by factors of two, then adding the impact of
those variations linearly.
The uncertainty in the electroweak $K$-factor is evaluated from the
effect of neglecting real boson emission, varying the electroweak scheme
definitions as implemented in {\sc Pythia} and {\sc Horace},
as well as of the effect of
higher order electroweak and $\mathcal{O}(\alpha\alpha_S)$ corrections.
For the electron channel, the QCD background estimate is subject to
an uncertainty derived from a comparison with different background estimate
methods (see discussion above).
For the muon channel, the QCD background uncertainty is negligible.

Experimental uncertainties originate from
the energy/momentum resolution, as well as the trigger, reconstruction 
and identification efficiencies.
In the electron channel, the uncertainty in the constant
term, which dominates the energy resolution at high energy,
has a negligible impact on the analysis.
Knowledge of the energy scale also has a negligible effect.
The electron reconstruction and identification uncertainty results in
a 1.5\% effect, which is estimated by studying the impact of the isolation
requirement on the dielectron mass distribution.
In the muon channel, the momentum resolution is dominated by the quality
of the muon spectrometer alignment. The uncertainty in the alignment is
evaluated directly from dedicated toroid field-off runs and redundant
momentum measurements in overlapping small and large chambers.
These experimental uncertainties are found to have minimal impact
on the dimuon mass distribution.
Finally, a systematic error growing from 0.3\% at the $Z$ pole
to 4.5\% at 1.5~\TeV{} is assigned to the muon reconstruction
efficiency to account conservatively for its small $p_T$ dependence due to
occasional large energy loss from bremsstrahlung.

\section{\label{sec:analysis}Statistical Analysis}
The data analysis proceeds with a Bayesian method to compare the observed
event yields with the expected yields for a range of different contact interaction
model parameters.
Specifically, the number $\mu$ of expected events in each of the mass bins
defined in Tables~\ref{dielec_yields} and \ref{dimuon_yields}
is 
\begin{equation}
\mu = n_{\rm DY+CI}(\theta,\bar{\nu}) + n_{\rm non-DY~bg}(\bar{\nu})~,
\end{equation}
where $n_{\rm DY+CI}(\theta,\bar{\nu})$ is the number of events predicted
by the {\sc Pythia} DY+CI MC for a particular choice of contact interaction model 
parameter $\theta$, $n_{\rm non-DY~bg}(\bar{\nu})$ is the number of non-DY background
events,
and $\bar{\nu}$ represents the set of Gaussian nuisance parameters that account
for systematic uncertainties in these numbers as discussed above.
The parameter $\theta$ corresponds to a choice of energy scale $\Lambda$ and
interference parameter $\eta_{LL}$.
The complete set of $\mu$ values used in this analysis is shown 
in Tables~\ref{lambda_yields_elec} and \ref{lambda_yields_muon} 
for the electron and muon channels, respectively.
For each mass bin, a second order polynomial is used to model the dependence
of $\mu$ on $1/\Lambda^2$.

\begin{table*}[tb]
\newcommand\T{\rule{0pt}{2.6ex}}
\newcommand\B{\rule[-1.2ex]{0pt}{0pt}}
\caption{Expected numbers of events in the signal region of the analysis
  for various contact interaction scales with constructive ($\Lambda^-$)
  and destructive ($\Lambda^+$) interference in the electron channel. 
  The errors quoted originate from both systematic uncertainties and limited MC statistics.}
\label{lambda_yields_elec}
\centering
\scriptsize
\begin{tabular}{c r@{\ $\pm$\ }l r@{\ $\pm$\ }l r@{\ $\pm$\ }l r@{\ $\pm$\ }l r@{\ $\pm$\ }l}
\hline
\hline
$m_{ee}$ [\GeV{}] \T \B & \multicolumn{2}{c}{150-170} & \multicolumn{2}{c}{170-200} & \multicolumn{2}{c}{200-240} & \multicolumn{2}{c}{240-300} & \multicolumn{2}{c}{300-400} \\
\hline
$\Lambda^-$~=~3 \TeV{} \T & 785 & 29 & 649 & 26 & 467 & 22 & 383 & 19 & 343 & 12 \\
$\Lambda^-$~=~4 \TeV{} \T & 781 & 28 & 647 & 26 & 437 & 21 & 326 & 17 & 223 & 7 \\
$\Lambda^-$~=~5 \TeV{} \T & 734 & 27 & 612 & 24 & 405 & 19 & 298 & 16 & 181 & 6 \\
$\Lambda^-$~=~7 \TeV{} \T & 691 & 26 & 638 & 25 & 406 & 19 & 259 & 15 & 163 & 5 \\
$\Lambda^-$~=~12 \TeV{} \T \B & 721 & 26 & 604 & 24 & 336 & 17 & 234 & 14 & 149 & 5 \\
\hline
$\Lambda^+$~=~3 \TeV{} \T & 770 & 28 & 642 & 24 & 424 & 20 & 331 & 17 & 269 & 9 \\
$\Lambda^+$~=~4 \TeV{} \T & 745 & 27 & 591 & 23 & 385 & 19 & 277 & 16 & 166 & 5 \\
$\Lambda^+$~=~5 \TeV{} \T & 702 & 25 & 607 & 23 & 350 & 17 & 258 & 15 & 151 & 5  \\
$\Lambda^+$~=~7 \TeV{} \T & 672 & 25 & 600 & 23 & 399 & 19 & 251 & 14 & 142 & 5 \\
$\Lambda^+$~=~12 \TeV{} \T \B & 749 & 27 & 593 & 23 & 403 & 19 & 274 & 15 & 137.9 & 4.4  \\
\hline
\hline
 \multicolumn{11}{c}{ }  \\
\hline
\hline
$m_{ee}$ [\GeV{}] \T \B & \multicolumn{2}{c}{400-550} & \multicolumn{2}{c}{550-800} & \multicolumn{2}{c}{800-1200} & \multicolumn{2}{c}{1200-1800} & \multicolumn{2}{c}{1800-3000} \\
\hline
$\Lambda^-$~=~3 \TeV{} \T & 286 & 11 & 269 & 12  & 207 & 11 & 112 & 8 & 30.3 & 3.4 \\
$\Lambda^-$~=~4 \TeV{} \T & 132 & 5 & 109.1 & 4.9 &  80.5 & 4.1 &  29.8 & 2.3 & 10.9 & 1.3\\
$\Lambda^-$~=~5 \TeV{} \T & 82.7 & 3.6 & 57.3 & 2.6 & 35.5 & 1.9 & 14.8 & 1.1 & 3.9 & 0.5 \\
$\Lambda^-$~=~7 \TeV{} \T & 68.3 & 2.9 & 29.0 & 1.4 & 11.2 & 0.7 & 4.0 & 0.4 & 1.08 & 0.17 \\
$\Lambda^-$~=~12 \TeV{} \T \B & 57.3 & 2.6 & 18.8 & 1.1 & 5.0 & 0.4 & 1.00 & 0.13 & 0.15 & 0.05 \\
\hline
$\Lambda^+$~=~3 \TeV{} \T & 215 & 8 & 239 & 11 & 185 & 10 & 107 & 7 & 28.4 & 3.2 \\
$\Lambda^+$~=~4 \TeV{} \T & 100.9 & 3.8 & 78.1 & 3.6 & 60.7 & 3.2 & 31.5 & 2.2 & 8.7 & 1.0 \\
$\Lambda^+$~=~5 \TeV{} \T & 64.4 & 2.7 & 36.2 & 1.8 & 25.4 & 1.3 & 12.7 & 0.9 & 3.5 & 0.4 \\
$\Lambda^+$~=~7 \TeV{} \T & 56.3 & 2.5 & 20.3 & 1.1 & 7.7 & 0.5 & 3.58 & 0.28 & 0.83 & 0.12 \\
$\Lambda^+$~=~12 \TeV{} \T \B &  52.4 & 2.4 &  14.4 & 0.9 &  3.29 & 0.28 &  0.46 & 0.08 & 0.075 & 0.026 \\
\hline
\hline
\end{tabular}
\end{table*}

\begin{table*}[tb]
\newcommand\T{\rule{0pt}{2.6ex}}
\newcommand\B{\rule[-1.2ex]{0pt}{0pt}}
\caption{Expected numbers of events in the signal region of the analysis
  for various contact interaction scales with constructive ($\Lambda^-$)
  and destructive ($\Lambda^+$) interference in the muon channel.
  The errors quoted originate from both systematic uncertainties and limited MC statistics.}
\label{lambda_yields_muon}
\centering
\scriptsize
\begin{tabular}{c r@{\ $\pm$\ }l r@{\ $\pm$\ }l r@{\ $\pm$\ }l r@{\ $\pm$\ }l r@{\ $\pm$\ }l}
\hline
\hline
$m_{\mu\mu}$ [\GeV{}] \T \B & \multicolumn{2}{c}{150-170} & \multicolumn{2}{c}{170-200} & \multicolumn{2}{c}{200-240} & \multicolumn{2}{c}{240-300} & \multicolumn{2}{c}{300-400} \\
\hline
$\Lambda^-$~=~3 \TeV{} \T & 638 & 28 & 547 & 26 & 371 & 22 & 285 & 19 & 263 & 13\\
$\Lambda^-$~=~4 \TeV{} \T & 618 & 27 & 513 & 24 & 287 & 18 & 228 & 15 & 163 & 7\\
$\Lambda^-$~=~5 \TeV{} \T & 572 & 26 & 478 & 23 & 357 & 20 & 206 & 14 & 131 & 6\\
$\Lambda^-$~=~7 \TeV{} \T & 571 & 25 & 496 & 23 & 294 & 18 & 187 & 14 & 113 & 5\\
$\Lambda^-$~=~12 \TeV{} \T \B & 606 & 26 & 441 & 22 & 252 & 16 & 191 & 14 & 100.0 & 3.8\\
\hline
$\Lambda^+$~=~3 \TeV{} \T & 602 & 26 & 417 & 21 & 332 & 19 & 205 & 15 & 186 & 10\\
$\Lambda^+$~=~4 \TeV{} \T & 575 & 25 & 456 & 22 & 286 & 17 & 182 & 13 & 112 & 5\\
$\Lambda^+$~=~5 \TeV{} \T & 554 & 25 & 483 & 23 & 289 & 17 & 167 & 12 & 102.0 & 4.0\\
$\Lambda^+$~=~7 \TeV{} \T & 557 & 24 & 435 & 21 & 292 & 18 & 196 & 14 & 102.0 & 4.4\\
$\Lambda^+$~=~12 \TeV{} \T \B & 576 & 25 & 421 & 21 & 256 & 16 & 186 & 13 & 100.0 & 3.9 \\
\hline
\hline
 \multicolumn{11}{c}{ }  \\
\hline
\hline
$m_{\mu\mu}$ [\GeV{}] \T \B & \multicolumn{2}{c}{400-550} & \multicolumn{2}{c}{550-800} & \multicolumn{2}{c}{800-1200} & \multicolumn{2}{c}{1200-1800} & \multicolumn{2}{c}{1800-3000} \\
\hline
$\Lambda^-$~=~3 \TeV{} \T & 192 & 11 & 185 & 12 & 151 & 12 & 60 & 9 & 18 & 7 \\
$\Lambda^-$~=~4 \TeV{} \T & 97 & 5 & 73 & 5 & 50.3 & 4.0 & 17.9 & 2.9 & 6.8 & 2.4 \\
$\Lambda^-$~=~5 \TeV{} \T & 62.8 & 3.3 & 37.5 & 2.2 & 21.8 & 1.8 & 8.7 & 1.3 & 2.8 & 1.0 \\
$\Lambda^-$~=~7 \TeV{} \T & 45.9 & 2.4 & 19.5 & 1.2 & 7.9 & 0.7 & 3.0 & 0.5 & 0.59 & 0.38 \\
$\Lambda^-$~=~12 \TeV{} \T \B & 38.0 & 2.0 & 14.0 & 0.9 & 2.90 & 0.26 & 0.91 & 0.24 & 0.11 & 0.14 \\
\hline
$\Lambda^+$~=~3 \TeV{} \T & 152 & 9 & 153 & 11 & 131 & 11 & 63 & 8 & 19 & 6 \\
$\Lambda^+$~=~4 \TeV{} \T & 68.2 & 3.7 & 54.7 & 3.7 & 42.8 & 3.4 & 22.2 & 2.7 & 7.4 & 2.1 \\
$\Lambda^+$~=~5 \TeV{} \T & 51.0 & 2.5 & 24.9 & 1.6 & 15.7 & 1.3 & 8.0 & 1.0 & 2.9 & 0.8 \\
$\Lambda^+$~=~7 \TeV{} \T & 33.9 & 1.8 & 13.3 & 0.8 & 5.04 & 0.43 & 1.85 & 0.32 & 0.63 & 0.30 \\
$\Lambda^+$~=~12 \TeV{} \T \B & 34.0 & 1.8 & 10.1 & 0.6 & 1.73 & 0.18 & 0.25 & 0.12 & 0.07 & 0.10 \\
\hline
\hline
\end{tabular}
\end{table*}

The likelihood of observing a set of $\bar{n}$ events in $N$ invariant mass bins
is given by a product of Poisson probabilities for each mass bin $k$:
\begin{equation}\label{likelihood}
 \text{$\cal L$} (\bar{n} \mid \theta,\bar{\nu}) = \displaystyle\prod_{k=1}^{N}\frac{\mu_k^{n_k}e^{-\mu_k}}{n_k!}.
\end{equation}
According to Bayes' theorem, the posterior probability
for the parameter $\theta$ given $\bar{n}$ observed events
is
\begin{equation}\label{posterior}
 \text{$\cal P$} (\theta \mid \bar{n}) = \frac{1}{\text{$\cal Z$}} \text{$\cal L_M$} (\bar{n} \mid \theta) P(\theta),
\end{equation}
where ${\cal Z}$ is a normalization constant and
the marginalized likelihood $\cal L_M$ corresponds to the likelihood after
all nuisance parameters have been integrated out.
This integration is performed assuming that the nuisance parameters are correlated
across all mass bins and that they affect both signal and background expectations,
except for the electroweak $K$-factor that only affects the DY and DY+CI components.
The prior probability $P(\theta)$ is chosen to be flat in $1/\Lambda^2$,
motivated by the form of Eq.~(\ref{cross_section}).
The 95\% CL limit is then obtained by finding the value 
$\Lambda_{\rm lim}$ satisfying
$\int_0^{\theta_{\rm lim}} \text{$\cal P$} (\theta \mid \bar{n}) \: d\theta = 0.95$,
where $\theta = 1/\Lambda^2$.
The above calculations have been performed with the Bayesian Analysis Toolkit (BAT)~\cite{BAT}.

\section{\label{sec:results}Results}
To test the consistency between the data and the standard model,
a likelihood ratio test was performed by producing a set of 1000
SM-like pseudoexperiments and comparing the likelihood ratio
between the signal+background and pure background hypotheses obtained
in the data to the results of the pseudoexperiments.
The signal+background likelihood is evaluated 
at the $\Lambda$ value that maximizes it.
The derived {\it p-value}, corresponding to the probability of observing
a fluctuation in the pseudoexperiments
that is at least as signal-like as that seen in the data
(i.e. with a maximum likelihood ratio greater or equal to that obtained in the data),
is estimated to be 39\% (79\%) in the electron channel
and 21\% (5\%) in the muon channel for constructive (destructive) interference.
These values indicate that there is no significant evidence for
contact interactions in the analyzed data and thus limits are
set on the contact interaction scale $\Lambda$.

Using the Bayesian method described above,
the expected 95\% CL lower limit values on the energy scale $\Lambda$ are found to be
$\expLimitMinusElec \pm 1.0$ \TeV{} for constructive interference ($\eta_{LL} = -1$)
and $\expLimitPlusElec \pm 1.0$ \TeV{} for destructive interference ($\eta_{LL} = +1$)
in the electron channel.
The corresponding expected limits in the muon channel are
$\expLimitMinusMuon \pm 0.9$ \TeV{} and $\expLimitPlusMuon \pm 0.9$ \TeV{}.
The quoted uncertainties correspond to the 68\% range of limits surrounding the
median value of all limits obtained with a set of 1000 pseudoexperiments.
Stronger limits are expected in the electron channel than in the muon channel
due to the significantly larger acceptance for the dielectron selection.

The observed limits (at 95\% CL) are
$\Lambda^- > \obsLimitMinusElec$~\TeV{} ($\Lambda^+ > \obsLimitPlusElec$~\TeV{})
in the electron channel
and
$\Lambda^- > \obsLimitMinusMuon$~\TeV{} ($\Lambda^+ > \obsLimitPlusMuon$~\TeV{})
in the muon channel
for constructive (destructive) interference.
These limits are summarized in Table~\ref{limits}.

If instead of choosing the prior to be flat in $1/\Lambda^2$,
it is selected to be flat in $1/\Lambda^4$ to match the form of the
pure CI term in Eq.~(\ref{cross_section}), the observed limit in the
electron channel becomes weaker
by 0.9~\TeV{} (0.8~\TeV{}) for constructive (destructive) interference.
The corresponding shift to lower values is 0.4~\TeV{} (0.3~\TeV{}) in the muon channel,
see Table~\ref{limits}.

The quoted limits have been obtained with electroweak corrections applied
to both DY and DY+CI samples for consistency, 
although part of the electroweak corrections cannot
be computed reliably due to the unknown new physics represented by the
contact interaction. However, this particular choice does not affect the observed
limits significantly and results in slightly more conservative limits. 
The limits obtained without electroweak corrections
improve by less than 0.1~\TeV{}.

Finally, a limit is set on the combination of the electron and muon channels, 
assuming lepton universality, by computing a combined posterior probability 
for the two channels. The following sources of systematic
uncertainty are treated as fully correlated between the two channels: 
PDF and $\alpha_S$, QCD and electroweak $K$-factors,
and $Z/\gamma^\ast$ cross section for normalization.
The resulting combined limits are
$\Lambda^- > \obsLimitMinusComb$~\TeV{} and $\Lambda^+ > \obsLimitPlusComb$~\TeV{}
for the $1/\Lambda^2$ prior.
Table~\ref{limits} summarizes all limits for the two priors considered in
this study and shows
that the combined limit is weaker than the electron channel limit
in the case of destructive interference. This is due to the slight excess of
events observed in the muon channel for masses greater than 800 \GeV{},
an excess that gives an increase in the likelihood that is stronger for
destructive interference than it is for constructive interference.

\begin{table}[htb]
\newcommand\T{\rule{0pt}{2.6ex}}
\newcommand\B{\rule[-1.2ex]{0pt}{0pt}}
  \begin{center}
    \caption{Expected and observed 95\% CL lower limits on the contact interaction
     energy scale $\Lambda$ for the electron and muon channels, as well as for
     the combination of those channels. Separate results are provided for
     the different choices of flat priors: $1/\Lambda^2$ and $1/\Lambda^4$.\label{limits}}
    \begin{tabular}{lccccc}
      \hline\hline
      Channel \T & Prior & \multicolumn{2}{c}{Expected limit (\TeV{})} & \multicolumn{2}{c}{Observed limit (\TeV{})} \\
              &       & Constr. & Destr. & Constr. & Destr. \\
      \hline
      $e^+ e^-$ \T & $1/\Lambda^2$ & $\expLimitMinusElec$ & $\expLimitPlusElec$ & $\obsLimitMinusElec$ & $\obsLimitPlusElec$ \\
              \B  & $1/\Lambda^4$ & 8.9 & 8.6 & 9.2 & 8.6 \\ \hline
      $\mu^+\mu^-$ \T & $1/\Lambda^2$ & $\expLimitMinusMuon$ & $\expLimitPlusMuon$ & $\obsLimitMinusMuon$ & $\obsLimitPlusMuon$ \\
              \B  & $1/\Lambda^4$ & 8.3 & 7.9 & 7.6 & 6.7 \\ \hline
      Comb. \T & $1/\Lambda^2$ & $\expLimitMinusComb$ & $\expLimitPlusComb$ & $\obsLimitMinusComb$ & $\obsLimitPlusComb$ \\
              \B & $1/\Lambda^4$ & 9.6 & 9.4 & 9.4 & 8.4 \\
      \hline\hline
    \end{tabular}
  \end{center}
\end{table}

\section{\label{sec:conclusions}Conclusions}
A search for contact interactions in $e^+ e^-$ and $\mu^+ \mu^-$ events
produced in proton-proton collisions at $\sqrt{s} = 7$~\TeV{} has been
performed.
The analysis uses early 2011 run data amounting to
$\integLumiElec$ ($\integLumiMuon$) fb$^{-1}$ of $pp$ collisions
in the electron (muon) channel collected with the ATLAS detector.
The dilepton mass distributions do not display significant deviations
from the standard model.
Using a Bayesian approach with a $1/\Lambda^2$ prior, as was
done in most previous searches at hadron colliders,
the following 95\% CL limits are set on the energy scale of contact interactions:
$\Lambda^- > \obsLimitMinusElec$~\TeV{} ($\Lambda^+ > \obsLimitPlusElec$~\TeV{})
in the electron channel and
$\Lambda^- > \obsLimitMinusMuon$~\TeV{} ($\Lambda^+ > \obsLimitPlusMuon$~\TeV{})
in the muon channel
for constructive (destructive) interference in the left-left isoscalar
compositeness model.
Somewhat weaker limits are obtained with a prior flat in $1/\Lambda^4$.
These limits are the most stringent to date on $\mu\mu qq$ contact interactions
and exceed the best existing limits set by a single
experiment on $e e q q$ contact interactions for light-quark flavors.

\section*{Acknowledgements}

We thank CERN for the very successful operation of the LHC, as well as the
support staff from our institutions without whom ATLAS could not be
operated efficiently.

We acknowledge the support of ANPCyT, Argentina; YerPhI, Armenia; ARC,
Australia; BMWF, Austria; ANAS, Azerbaijan; SSTC, Belarus; CNPq and FAPESP,
Brazil; NSERC, NRC and CFI, Canada; CERN; CONICYT, Chile; CAS, MOST and
NSFC, China; COLCIENCIAS, Colombia; MSMT CR, MPO CR and VSC CR, Czech
Republic; DNRF, DNSRC and Lundbeck Foundation, Denmark; ARTEMIS, European
Union; IN2P3-CNRS, CEA-DSM/IRFU, France; GNAS, Georgia; BMBF, DFG, HGF, MPG
and AvH Foundation, Germany; GSRT, Greece; ISF, MINERVA, GIF, DIP and
Benoziyo Center, Israel; INFN, Italy; MEXT and JSPS, Japan; CNRST, Morocco;
FOM and NWO, Netherlands; RCN, Norway; MNiSW, Poland; GRICES and FCT,
Portugal; MERYS (MECTS), Romania; MES of Russia and ROSATOM, Russian
Federation; JINR; MSTD, Serbia; MSSR, Slovakia; ARRS and MVZT, Slovenia;
DST/NRF, South Africa; MICINN, Spain; SRC and Wallenberg Foundation,
Sweden; SER, SNSF and Cantons of Bern and Geneva, Switzerland; NSC, Taiwan;
TAEK, Turkey; STFC, the Royal Society and Leverhulme Trust, United Kingdom;
DOE and NSF, United States of America.

The crucial computing support from all WLCG partners is acknowledged
gratefully, in particular from CERN and the ATLAS Tier-1 facilities at
TRIUMF (Canada), NDGF (Denmark, Norway, Sweden), CC-IN2P3 (France),
KIT/GridKA (Germany), INFN-CNAF (Italy), NL-T1 (Netherlands), PIC (Spain),
ASGC (Taiwan), RAL (UK) and BNL (USA) and in the Tier-2 facilities
worldwide.


\onecolumngrid
\clearpage
\input{atlas_authlist}

\end{document}

%% file: atlas_authlist.tex
\begin{flushleft}
{\Large The ATLAS Collaboration}

\bigskip

G.~Aad$^{\rm 48}$,
B.~Abbott$^{\rm 110}$,
J.~Abdallah$^{\rm 11}$,
A.A.~Abdelalim$^{\rm 49}$,
A.~Abdesselam$^{\rm 117}$,
O.~Abdinov$^{\rm 10}$,
B.~Abi$^{\rm 111}$,
M.~Abolins$^{\rm 87}$,
O.S.~AbouZeid$^{\rm 157}$,
H.~Abramowicz$^{\rm 152}$
% ...
\bigskip

% ...
$^{10}$ Institute of Physics, Azerbaijan Academy of Sciences, Baku, Azerbaijan\\
$^{11}$ Institut de F\'isica d'Altes Energies and Departament de F\'isica de la Universitat Aut\`onoma  de Barcelona and ICREA, Barcelona, Spain\\
% ...
$^{48}$ Fakult\"{a}t f\"{u}r Mathematik und Physik, Albert-Ludwigs-Universit\"{a}t, Freiburg i.Br., Germany\\
$^{49}$ Section de Physique, Universit\'e de Gen\`eve, Geneva, Switzerland\\
% ...
$^{87}$ Department of Physics and Astronomy, Michigan State University, East Lansing MI, United States of America\\
% ...
$^{110}$ Homer L. Dodge Department of Physics and Astronomy, University of Oklahoma, Norman OK, United States of America\\
$^{111}$ Department of Physics, Oklahoma State University, Stillwater OK, United States of America\\
% ...
$^{117}$ Department of Physics, Oxford University, Oxford, United Kingdom\\
% ...
$^{152}$ Raymond and Beverly Sackler School of Physics and Astronomy, Tel Aviv University, Tel Aviv, Israel\\
% ...
$^{157}$ Department of Physics, University of Toronto, Toronto ON, Canada\\
% ...
\end{flushleft}


%% file: ci_paper_v2.bbl
\begin{thebibliography}{99}
\bibitem{coordinates}
ATLAS uses a right-handed coordinate system with its origin at the nominal
interaction point (IP) in the center of the detector and the $z$-axis along
the beam pipe. The $x$-axis points from the IP to the center of the LHC ring,
and the $y$-axis points upward. Cylindrical coordinates $(r,\phi)$
are used in the transverse plane, $\phi$ being the azimuthal angle around
the beam pipe. The pseudorapidity is defined in terms of the polar
angle $\theta$ as $\eta$~=~$-$~$\ln\tan(\theta /2)$.
\end{thebibliography}
